\documentclass[11pt,dvips]{article}
cm \voffset = -2truecm

\usepackage{graphicx}
\usepackage{amssymb}
\usepackage{latexsym}
\usepackage{color}

\begin{document}
\begin{titlepage}
\setcounter{page}{1}
\renewcommand{\thefootnote}{\fnsymbol{footnote}}

\vspace{5mm}
\begin{center}

 {\Large \bf Quantum discord of Bell cat-states under amplitude damping}

\vspace{1.5cm}

{\bf M. Daoud}$^{a}${\footnote { email: {\sf
m$_{-}$daoud@hotmail.com}}} and {\bf R. Ahl Laamara}$^{b,c}$
{\footnote { email: {\sf ahllaamara@gmail.com}}}

\vspace{0.5cm}
$^{a}${\it Department of Physics, Faculty of Sciences, University Ibnou Zohr,\\
 Agadir,
Morocco}\\[1em]

$^{b}${\it LPHE-Modeling and Simulation, Faculty  of Sciences,
University
Mohammed V,\\ Rabat, Morocco}\\[1em]

$^{c}${\it Centre of Physics and Mathematics,
CPM, CNESTEN,\\ Rabat, Morocco}\\[1em]

\vspace{3cm}

\begin{abstract}

The evolution of pairwise quantum correlations of Bell cat-states
under amplitude damping is examined using the concept of quantum
discord which goes beyond entanglement. A closed expression of the
quantum discord is explicitly derived. We used of the Koashi-Winter
relation. A relation which facilitates the optimization process of
the conditional entropy. We also discuss the temporal evolution of
bipartite quantum correlations under a dephasing channel and compare
the behaviors of quantum discord and entanglement whose properties
are characterized through the concurrence.

\end{abstract}
\end{center}
\end{titlepage}

\newpage

\section{Introduction}

Quantum entanglement plays an important role in the area of quantum
information such as quantum teleportation \cite{Ben1}, superdense
coding \cite{Ben2}, quantum key distribution \cite{Eckert},
telecloning \cite{Murao} and many more. It is one of the most
fundamental concepts in the description of quantum correlations in
multipartite quantum systems and  makes possible tasks in quantum
information which are impossible without it. Therefore it was
important to quantify the amount of quantum correlations in a given
bipartite quantum state. Much effort has been devoted to the
classification of quantum states into separable and entangled states
(see \cite{NC-QIQC-2000,Alber-QI2001, Horodecki-RMP-2009} and
references therein). Until some time ago, entanglement was usually
regarded as the only kind of nonclassical correlation in a composed
state. However, there are other nonclassical correlations different
from those involved in entanglement which are useful for quantum
technology. Hence, to characterize all nonclassical correlations
present in a multipartite system, the so-called quantum discord,
which goes beyond entanglement, was introduced in
\cite{Vedral-et-al,Ollivier-PRL88-2001}. Now, it is commonly
accepted that quantum discord constitutes a new resource for quantum
computation. Among the evidences of the relevance of the quantum
discord, one may quote for instance quantum non-locality without
entanglement \cite{Bennett-PRA59-1999, Horodecki-PRA71-2005,
Niset-PRA74-2006} and the advantages offered in increasing the
rapidity  of certain computational tasks with separable states in
comparison with their classical counterparts
\cite{Braunstein-PRL83-1999, Meyer-PRL85-2000,
Datta-PRL100-2008,Lanyon-PRL101-2008}. But, it may be noticed that
despite increasing evidences for relevance of the quantum discord in
information processing tasks, there is no straightforward method to
get the analytical form of quantum discord in a given quantum state.
This is mainly due to the fact that its evaluation involves an
optimization procedure which is in general a hard task to perform. A
reliable algorithm to evaluate quantum discord for general two-qubit
states is still missing and only few analytical results were
obtained for some particular forms of the so-called two-qubit
X-states \cite{Luo,Ali,Shi1,Girolami,Shi2,Fanchini}. This class
includes the maximally entangled Bell states and Werner states
\cite{Wer-PRA89}. A closed expression for the discord of arbitrary
states remains an important challenge.

On the other hand, quantum optical tools are expected to be useful
 in the context of quantum information science, especially for
 communications using qubits over long distance. However, optical
 qubits suffer from decoherence due to energy loss or photon
 absorption. To reduce the decoherence effects, encoding qubits
 in multi-photon  optical coherent states seems to be a promising alternative.
 They are more robust against small levels of photon absorption (see
 \cite{Glancy}). The photon loss or amplitude damping in a noisy
 environment can be modeled by assuming that some of field energy and
 information is lost after transmission through a beam splitter.

Considering this problem and motivated by many works devoted to
study entanglement properties in a composite system involving
coherent states (for a recent review see \cite{Sanders}), we
investigate in this paper a method to  describe the evolution of
quantum discord in Bell cat-states under amplitude damping. We
present an algorithm to calculate the quantum discord. First, we
obtain an explicit and simplified expression for the conditional
entropy, exploiting the Bloch representation of the density matrix.
Then, we combine the purification method and the Koashi-Winter
relation \cite{Koachi-Winter} to perform easily the optimization of
the conditional entropy. This allows us to get a closed form of
quantum discord in damped Bell cat-states. Exploiting our algorithm,
we examine the dynamical evolution of the system under a dephasing
channel to compare the temporal behaviors of quantum discord and
entanglement as quantifiers of non-classical correlations.

This paper is organized as follows. Section 2 provides an
introduction to quantum discord. We present the main definitions and
properties. Section 3 concerns Glauber coherent states
superpositions subjected to an amplitude damping channel. In the
optical context, amplitude damping can be appropriately modeled by
having the signal interact with a vacuum mode in a beam splitter.
Having identified the effect of amplitude damping, we study in
section 4 the evolution of the quantum correlations of Bell
cat-states under amplitude damping. We obtain the explicit form of
quantum discord. Section 5 deals with the comparison of the
evolution of quantum discord and entanglement under a dephasing
channel, using the results of the previous sections. Finally, the
last section recalls the main results of our work and suggests
further issues deserving  to be investigated.

\section{Quantum discord: Generalities}

For a state $\rho^{AB}$  of a bipartite quantum system composed of
particle $A$ and particle $B$, the quantum discord is defined as the
difference between total correlation $I(\rho^{AB})$ and classical
correlation $C(\rho^{AB})$. The total correlation is usually
quantified by the mutual information $I$
\begin{equation}
    I(\rho^{AB})=S(\rho^A)+S(\rho^B)-S(\rho^{AB}),
\end{equation}
where $\rho^{A(B)}={\rm Tr}_{B(A)}(\rho^{AB})$ is the reduced state
of $A$($B$), and $S(\rho)$ is the von Neumann entropy of a quantum
state $\rho$. Suppose that a positive operator valued measure (POVM)
measurement is performed on particle $A$. The set of POVM elements
is denoted by $\mathcal{M}=\{M_k\}$ with $M_k\geqslant 0$ and
$\sum_k M_k= \mathbb{I} $. We remind that the generalized positive
operator valued measurement is not required. Indeed, it has be shown
in \cite{Hamieh} that for the optimal measurement for the
conditional entropy is ensured by projective one. Thus, a projective
measurement on the subsystem $A$ project the system into a
statistical ensemble $\{ p^B_k , \rho^B_{k}\}$, such that
\begin{eqnarray}
\rho^{AB} \longrightarrow \rho^B_{k} = \frac{(M_k \otimes
\mathbb{I})\rho^{AB}(M_k \otimes \mathbb{I})}{p^B_k}\label{rhoBk}
\end{eqnarray}
where the von Neumann measurement for subsystem $A$ writes as
\begin{eqnarray}
M_k = U \, \Pi_k \, U^\dagger : \quad k = 0,1 \, , \label{Eq:VNmsur}
\end{eqnarray}
with $\Pi_k = |k\rangle\langle k|$ is the projector for subsystem
$A$  along the computational base $|k\rangle$,  $U \in SU(2)$ is a
unitary operator with unit determinant, and
$$ p^B_k = {\rm tr}  \bigg[ (M_k \otimes \mathbb{I})\rho^{AB}(M_k \otimes \mathbb{I}) \bigg].$$
The amount of information acquired about particle $B$ is then given
by
$$S(\rho^B)-\sum_k ~p^B_k ~S(\rho^B_k),$$
which depends on measurement $\mathcal{M}$. This dependence can be
removed by doing maximization over all the measurements, which gives
rise to the definition of classical correlation:
\begin{eqnarray}
    C(\rho^{AB})& =\max_{\mathcal{M}}
    \Big[S(\rho^B)-\sum_k ~p^B_k ~S(\rho^B_k)\Big] \nonumber \\
    & =S(\rho^B) - \widetilde{S}_{\rm min}
      \label{def: classical correlation}
\end{eqnarray}
where $\widetilde{S}_{\rm min}$  denotes the minimal value of the
conditional  entropy
\begin{equation}
\widetilde{S} =  \sum_k ~p^B_k ~S(\rho^B_k).\label{condit-entropy}
\end{equation}
Then, the difference between $I(\rho^{AB})$ and $C(\rho^{AB})$ gives
the amount of quantum discord in the system
\begin{equation} \label{def: discord}
    D(\rho^{AB})= I(\rho^{AB}) - C(\rho^{AB})
    =S(\rho^A)+\widetilde{S}_{\rm min}-S(\rho^{AB}).
\end{equation}
The main difficulty, for several qubits as well as qudits systems,
lies in performing the minimization of the conditional entropy. This
explains why  there is no straightforward algorithm to compute
explicitly  quantum discord for mixed states. Only
 partial results are available. They were obtained for some special forms of the so-called
 X-states \cite{Li-Luo-PRA-2008,Dillenschneider-PRB78-2008, Sarandy-arXiv,Werlang-PRA-2009}.

\section{ Amplitude damping}
The beam splitter offers a simple way to probe the quantum nature of
electromagnetic field through simple experiments. The study of
entangled states  has revived interest in this device. Many authors
have considered the behavior of quantum states when passed through a
beam splitter \cite{Tan,Paris}. Recently, a quantum network of beam
splitters was used to create multi-particle entangled states of
continuous variables \cite{Van} and also multi-particle entangled
coherent states \cite{Wang2}. It also provides, as mentioned in the
introduction, a simple way to model the amplitude damping related to
the absorption of transmitted photons in a noisy channel.

\subsection{ Fock state inputs}

The beam splitter is an optical device with two input and two output
ports that, in some sense, governs the interaction  of two harmonic
oscillators. The input and output boson operators are related by a
unitary transformation which is an element of the $SU(2)$ group
defined by

\begin{equation}
{\cal B}(\theta) = \exp\left[\frac{\theta}{2}  \left(a^-_1 a^+_2 -
a^+_1 a^-_2\right)\right].
\end{equation}
 The objects $a^+_l$ and $a^-_l$ $( l = 1 , 2)$ are the usual harmonic
oscillator ladder operators. The reflection and transmission
coefficients
\begin{equation}
 t = \cos\frac{\theta}{2}~, \qquad  r = \sin\frac{\theta}{2}
\end{equation}
are defined  in terms of the angle $\theta$. The operator ${\cal B}$
is actually acting on the states $\vert n_1 , n_2 \rangle$ of the
usual  two dimensional harmonic oscillator. If the input state is
$\vert n_1 , n_2 \rangle$, then the ${\cal B} $ action leads to the
following Fock states superposition
\begin{equation}
{\cal B} \vert n_1 , n_2 \rangle = \sum_{m_1,m_2}  \langle m_1 , m_2
\vert {\cal B}
 \vert n_1 , n_2 \rangle  \vert m_1 , m_2 \rangle= \sum_{m_1,m_2} {\cal B}_{n_1,
n_2}^{m_1, m_2}
 \vert m_1 , m_2 \rangle
\end{equation}
and in general the output is a two-particle entangled state. On the
other hand, the action of the unitary operator ${\cal B}$ on the
state $\vert n , 0 \rangle$ gives
\begin{eqnarray}
{\cal B} \vert n , 0 \rangle &=& \left(1 +  |\xi|^2
\right)^{-\frac{n}{2}} \sum_{m=0}^{n} \xi^{m}
\frac{\sqrt{n!}}{\sqrt{(n - m)!m!}}  \vert n - m , m \rangle
\label{suncs}
\end{eqnarray}
where  the new variable $\xi =  r/t$ is defined as the ratio of the
reflection and transmission coefficients of the beam splitter under
consideration. Then, the output state (\ref{suncs}) turns out to be
the $SU(2)$ coherent state associated with the unitary
representation labeled by the integer $n$. This method can be
extended to a chain of $k$  beam splitters   to generate $SU(k+1)$
coherent states labeled by the reflection and transmission
parameters (see for instance \cite{daoud1,daoud2} where similar
notations were used). It is important to stress that the generation
of coherent states using beam splitters  requires input radiation
state with fixed number of photons. The experimental production of
such interesting and highly non classical states has been
investigated during the last decade (see \cite{Brattke} and
references therein). Recently an important experimental advance was
reported by Hofheinz et al in \cite{Hofheinz}. They gave the first
experimental demonstration for generating photon number Fock states
containing up to $n = 6$ photons in a super-conducting quantum
circuit.

\subsection{coherent state inputs}

To describe the photon loss, we usually assume that some of the
coherent field is lost in transit via a beam splitter. The coherent
states enters one port of the beam splitter and the vacuum,
representing the environment, enters the second port. After
transmission some information encoded in the coherent states is
transferred and the remaining amount of information is lost to the
noisy channel. To find the final state after transmission, we should
first evaluate the action of the operator ${\cal B}(\theta)$ on the
transmitted state. Here we shall be interested in the superpositions
of the form
\begin{equation}
|Q_\alpha\rangle = \frac{1}{\sqrt{N(\alpha)}} ( a |-\alpha\rangle +
b |\alpha\rangle ) \quad
\end{equation}
where $|a|^2 + |b|^2 = 1$ and $N(\alpha)$ is a normalization factor
given by
$$N(\alpha) = 1 + e^{-2|\alpha|^2}(ab^*+a^*b).$$
So, we start by evaluating the beam splitter action on the bipartite
state $\vert \alpha , 0 \rangle$ where the Glauber coherent-state $
|\alpha \rangle$ is expressed in the Fock (number) basis as
\begin{equation}
|\alpha\rangle = e^{-\frac{|\alpha|^2}{2}} \sum_{n=0}^{\infty}
\frac{\alpha^n}{\sqrt{n!}}|n\rangle.
\end{equation}
This yields
\begin{equation}
{\cal B}(\theta)\vert \alpha , 0 \rangle = \vert \alpha t , \alpha r
\rangle. \label{action1}
\end{equation}
It follows that the action of the operator ${\cal B}(\theta)$ on the
state $|Q_\alpha\rangle$ gives
\begin{equation}
|Q \rangle_t = \frac{1}{\sqrt{N(\alpha)}} ( a |-\alpha t, -\alpha r
\rangle  + b |\alpha t , \alpha r \rangle ),
\end{equation}
a state describing the quantum field and loss modes (environment).
The final state is then obtained by performing a partial trace over
the loss mode $l$. We get
\begin{equation}
\rho =\sum_{n=0}^{\infty} {}_{l}\langle n|Q\rangle_{t}~
{}_{t}\langle Q|n\rangle_{l}.
\end{equation}
A straightforward calculation gives
\begin{eqnarray}
\rho  =  \frac {1+c}{2}\bigg[ \bigg(\frac{1}{\sqrt{N(\alpha)}}(a|-
\alpha t \rangle + b | \alpha t \rangle ) \bigg) \times {\rm
h.c.}\bigg] + \frac {1- c}{2}\bigg[ \bigg(
\frac{1}{\sqrt{N(\alpha)}}( a|- \alpha t\rangle - b |\alpha
t\rangle) \bigg) \bigg] \times {\rm h.c.}\bigg],
\end{eqnarray}
where ${\rm h.c.}$ stands for Hermitian conjugation and $c$ is the
the coherent states overlapping $c = e^{-2r^2\vert
\alpha\vert^{2}}$. The last equation can be also written as
\begin{eqnarray}
\rho  = \frac{N(\alpha t)}{N(\alpha)} \bigg[ \frac
{1}{2}(1+c)|Q_{\alpha t}\rangle\langle Q_{\alpha t}| + \frac
{1}{2}(1-c) Z |Q_{\alpha t}\rangle\langle Q_{\alpha t}| Z\bigg],
\end{eqnarray}
where $Z$ is the Pauli $Z$-operator defined by
$$Z |Q_{\alpha t}\rangle = Z \bigg[ \frac{1}{\sqrt{N(\alpha t)}} ( a |\alpha t
\rangle  + b |\alpha t \rangle )\bigg] = \frac{1}{\sqrt{N(\alpha
t)}} ( a |\alpha t \rangle  - b |\alpha t \rangle )$$ which produces
a phase flip in the qubit basis. This implies that the transmission
of quantum information encoded in optical coherent states suffers
from two main types of error: the reduction of the amplitude of the
coherent state and the phase flip generated by the application of
the Pauli $Z$-operator.
\subsection{Two mode coherent state inputs}
The above considerations can be extended to two mode coherent states
of the form
\begin{equation}
\vert \chi_{\alpha, \alpha}\rangle = \frac{1}{\sqrt{{\cal
N}_{\alpha}}}(\sqrt{\omega}~\vert \alpha, \alpha \rangle +
e^{i\theta}\sqrt{1- \omega} ~\vert - \alpha, - \alpha \rangle)
\end{equation}
where
$${\cal
N}_{\alpha}^2 =  1 + 2\sqrt{\omega (1-\omega)} \cos\theta e^{-4\vert
\alpha \vert^2}. $$ The action of a beam splitter on the state
$\vert \chi_{\alpha, \alpha}\rangle \otimes \vert 0 \rangle$ gives
\begin{equation}
\vert \chi \rangle_t =  \vert \chi_{\alpha, \alpha}\rangle \otimes
\vert 0 \rangle = \frac{1}{\sqrt{{\cal
N}_{\alpha}}}(\sqrt{\omega}~\vert \alpha, \alpha t, \alpha r \rangle
+ e^{i\theta}\sqrt{1- \omega} ~\vert - \alpha, - \alpha t , - \alpha
r \rangle).
\end{equation}
The trace over the lost modes gives the density
\begin{equation}
\rho = {\rm Tr}_l ~\vert \chi \rangle_t ~_t\langle \chi \vert.
\end{equation}
Here again, one can see that the density $\rho$ takes the following
compact form
\begin{eqnarray}
\rho  = \frac{{\cal N}_{\alpha t}}{{\cal N}_{\alpha}} \bigg[ \frac
{1}{2}(1+c)|\chi_{\alpha,\alpha t}\rangle\langle \chi_{\alpha,\alpha
t}| + \frac {1}{2}(1-c) Z |\chi_{\alpha,\alpha t}\rangle\langle
\chi_{\alpha,\alpha t}| Z\bigg].
\end{eqnarray}
The final state is mixed and the amplitude of the second mode is
reduced.

\section{Evolution of Bell cat-states correlations under amplitude damping}
The Bell states are very interesting in quantum optics and have been
used in the field of quantum teleportation and many others quantum
computing operations. The experimental generation can be realized by
sending a cat states of the form
$|\sqrt{2}\alpha\rangle+|-\sqrt{2}\alpha\rangle$ and the vacuum into
the two input ports of a 50/50 beam splitter ${\cal B}(\pi/4)$ (see
 equation (\ref{action1})) to get
\begin{equation}
\vert B_{\alpha, \alpha}\rangle = \frac{1}{\sqrt{N_{\alpha}}}(\vert
\alpha, \alpha \rangle +  \vert - \alpha, - \alpha
\rangle)\label{bell}
\end{equation}
where the normalization factor is defined by
$$ N_{\alpha} = 2 ( 1 + e^{-4\vert \alpha \vert^2}).$$
Obviously, the problem of generating Bell states is deeply dependent
on the availability of a source of cat states. For instance, they
can be produced by sending a coherent state into a nonlinear medium
exhibiting the Kerr effect \cite{Yurke}. They can be also generated
using a squeezing interaction, linear optical devices and photon
counters \cite{Ralph03,Song}. Recently, a promising new method to
produce cat states was proposed by Lund {\it et al.} in \cite{Lund}.
The production of cat states especially ones  of high amplitude or
mean number of photons remains an experimental challenge.
Considering the fast technical progress and the increasing number of
groups working in this field, we expect that the generation of cat
states (and Bell states) is a goal that is achievable in the near
future.  The available experimental results in the literature,
obtained with the present day technology, are encouraging.
Superpositions of weak coherent states with opposite phase,
resembling to a small "Schr\"odinger's cat" state (or
"Schr\"odinger's kitten"), were produced by photon subtraction from
squeezed vacuum \cite{Ourjoumtsev}. Also, the experimental
generation of arbitrarily large squeezed cat states, using homodyne
detection and photon number states (two photons) as resources was
reported in \cite{Ourjoumtsev}.  Very recently, creation of coherent
state superpositions, by subtracting up to three photons from a
pulse of squeezed vacuum light, is reported in \cite{Gerrits}. The
 mean photon number of  such coherent states produced by
three-photon subtraction is of 2:75.


\subsection{Bell cat-states under amplitude damping.}

Using the results of the previous section, it is simple to verify
that under the action of a beam splitter on the state (\ref{bell}),
the resultant density is
\begin{eqnarray}\label{density-main}
\rho^{AB} = \frac{N_{\alpha t}}{N_{\alpha}} \bigg[ \frac
{1}{2}(1+c)|B_{\alpha, \alpha t}\rangle\langle B_{\alpha, \alpha t}|
+ \frac {1}{2}(1-c) Z |B_{\alpha, \alpha t}\rangle\langle B_{\alpha,
\alpha t}| Z\bigg].
\end{eqnarray}
To study the quantum correlations in this state, a qubit mapping is
required. It can be defined as follows.  For the first mode $A$, we
introduce a two dimensional basis spanned by the vectors $\vert
u_{\alpha} \rangle$ and $\vert v_{\alpha} \rangle$ defined by
\begin{equation}
\vert \alpha \rangle = a_{\alpha} \vert u_{\alpha} \rangle +
b_{\alpha}\vert v_{\alpha} \rangle \qquad \vert -\alpha \rangle =
a_{\alpha} \vert u_{\alpha} \rangle - b_{\alpha}\vert v_{\alpha}
\rangle
\end{equation}
where
$$\vert a_{\alpha} \vert^2 + \vert b_{\alpha} \vert^2 = 1\qquad
\vert a_{\alpha} \vert^2 - \vert b_{\alpha} \vert^2 = \langle
-\alpha \vert \alpha\rangle.$$ To simplify our purpose, we take
$a_{\alpha}$ and $b_{\alpha}$ reals:
$$a_{\alpha} = \frac{\sqrt{1 + p}}{\sqrt{2}} \quad b_{\alpha} = \frac{\sqrt{1 - p}}{\sqrt{2}} \qquad {\rm with}
 \quad p = \langle -\alpha \vert \alpha\rangle = e^{-2\vert \alpha \vert^2}.$$
Similarly, for the second mode $B$,  a two-dimensional basis
generated by the vectors $\vert u_{\alpha t} \rangle$ and $\vert
v_{\alpha t} \rangle$  is defined as
\begin{equation}
\vert \alpha t \rangle = a_{\alpha t} \vert u_{\alpha t} \rangle +
b_{\alpha t}\vert v_{\alpha t} \rangle \qquad \vert -\alpha t
\rangle = a_{\alpha t} \vert u_{\alpha t} \rangle - b_{\alpha
t}\vert v_{\alpha t} \rangle
\end{equation}
where
$$a_{\alpha t} = \frac{\sqrt{1 + p^{t^2}}}{\sqrt{2}} \quad b_{\alpha t} = \frac{\sqrt{1 - p^{t^2}}}{\sqrt{2}}.$$
The density matrix $\rho^{AB}$ (\ref{density-main}) can be cast in
 the following matrix form
\begin{eqnarray}
\rho^{AB} =  \frac{2}{N_{\alpha}}\left(
\begin{array}{cccc}
(1+c)a^2_{\alpha}a^2_{\alpha t} & 0 & 0 & (1+c)a_{\alpha}a_{\alpha
t}b_{\alpha
}b_{\alpha t}\\
0 & (1-c)a^2_{\alpha }b^2_{\alpha t} & (1-c)a_{\alpha}a_{\alpha
t}b_{\alpha
}b_{\alpha t} & 0 \\
0 & (1-c)a_{\alpha }a_{\alpha t}b_{\alpha }b_{\alpha t} &
(1-c)b^2_{\alpha }a^2_{\alpha t}
& 0 \\
(1+c)a_{\alpha} a_{\alpha t}b_{\alpha }b_{\alpha t} & 0 & 0 &
(1+c)b^2_{\alpha }b^2_{\alpha t}
\end{array}
\right)  \label{Xform}
\end{eqnarray}
in the representation spanned by two-qubit product states
$$|1\rangle = |u_{\alpha } \rangle_A
\otimes |u_{\alpha t} \rangle_B  \quad |2\rangle = | u_{\alpha}
\rangle_A \otimes | v_{\alpha t}\rangle_B \quad |3\rangle = |
v_{\alpha }\rangle_A \otimes |u_{\alpha t}\rangle_B \quad |4\rangle
= |v_{\alpha }\rangle_A \otimes |v_{\alpha t}\rangle_B.$$ The visual
form of the obtained density (\ref{Xform}) resembles the letter $X$
and it is a special kind  of the so-called called X-states which
have been extensively discussed in the literature
\cite{YuEberly-2007,Rau-JPA09}. The density $\rho^{AB}$ also
rewrites, in the Bloch representation, as
\begin{eqnarray}
\rho^{AB} = \frac{1}{4} (\mathbb{I} \otimes \mathbb{I} + R_{30}
~\sigma_3 \otimes \mathbb{I}  + R_{03}~ \mathbb{I} \otimes \sigma_3
+ \sum_{i = 1}^{3} ~R_{ii}~ \sigma_i \otimes \sigma_i )
\end{eqnarray}
where the correlation matrix elements are given by
\begin{eqnarray}
R_{03} = \frac{p^{t^2} + p^{2- t^2}}{1+p^2} , \qquad R_{30} =
\frac{2p}{1+p^2},\label{R03+R30}
\end{eqnarray}
\begin{eqnarray}
R_{11} = \frac{\sqrt{(1-p^2)(1-p^{2t^2})}}{1+p^2} , \qquad R_{22} =
-p^{1-t^2} \frac{\sqrt{(1-p^2)(1-p^{2t^2})}}{1+p^2} , \qquad R_{33}
= \frac{p^{1+t^2} + p^{1- t^2}}{1+p^2}.\label{Rii}
\end{eqnarray}

\subsection{Quantum mutual information}

The density $\rho^{AB}$ is a two qubit state of rank two. The
corresponding non vanishing  eigenvalues  are given by
\begin{eqnarray}
\lambda_1 =  \frac{(1+ p^{r^2})(1+ p^{t^2+1})}{2 + 2p^2} \qquad
\lambda_2 =  \frac{(1 - p^{r^2})(1 - p^{t^2+1})}{2 + 2p^2}.
\label{lambda12}
\end{eqnarray}
It follows that the joint entropy is
\begin{eqnarray}
S(\rho^{AB}) =  - \lambda_1 \log_2\lambda_1 - \lambda_2 \log_2
\lambda_2. \label{SAB}
\end{eqnarray}
The quantum mutual information is given by
\begin{eqnarray}
I (\rho^{AB}) = S(\rho^A) + S(\rho^B) + \sum_{j=1,2} \lambda_j
\log_2 \lambda_j \label{IAB}
\end{eqnarray}
where $\rho^A$ and $\rho^B$ are the marginal states of $\rho^{AB}$,
and
\begin{eqnarray}
S(\rho^A) =  - \lambda^A_+ \log_2\lambda^A_+ - \lambda^A_- \log_2
\lambda^A_- \qquad S(\rho^B) =  - \lambda^B_+ \log_2\lambda^B_+ -
\lambda^B_- \log_2 \lambda^B_- \label{SA+SB}
\end{eqnarray}
with
$$ \lambda^A_{\pm} = \frac{(1\pm p)^2}{2 + 2 p^2}\qquad \lambda^B_{\pm} = \frac{(1\pm p^{t^2})(1\pm p^{r^2+1})}{2 + 2 p^2}.$$
Reporting (\ref{SA+SB}) into (\ref{IAB}), the quantum mutual
information reads
\begin{eqnarray}
I(\rho^{AB})=  H\bigg(\frac{(1+ p)^2}{2 + 2 p^2}\bigg) +
H\bigg(\frac{(1+ p^{t^2})(1+ p^{r^2+1})}{2 + 2 p^2}\bigg)-
H\bigg(\frac{(1 + p^{r^2})(1 + p^{t^2+1})}{2 + 2p^2}\bigg)
\end{eqnarray}
where $H(x) = - x \log_2 x - (1-x) \log_2 (1 -x)$.

\subsection{Conditional entropy}
After computing the quantum mutual information, we need next to
compute the classical correlation $C(\rho^{AB})$ defined by
(\ref{def: classical correlation}). We consider projective
measurements for subsystem $A$. We follow the procedure developed in
\cite{Luo08}. We remind that the generalized positive operator
valued measurement is not required. Indeed, as mentioned above, the
optimal measurement for the conditional entropy is ensured by
projective operator \cite{Hamieh}. The general form of the $SU(2)$
unitary operator, occurring in (\ref{Eq:VNmsur}), is
\begin{eqnarray}
 U = \exp(\eta \sigma_+ - \bar\eta \sigma_-) \exp(i\phi
 \sigma_3)\label{operatorU}
\end{eqnarray}
where $\eta \in \mathbb{C}$ and $\phi \in \mathbb{R}$. This
parametrization allows us to express  the quantities defined by
$$\langle \sigma_i \rangle_ k = \langle k \vert U^\dagger \sigma_i  U \vert k \rangle, \qquad i = 1,2, 3 \quad {\rm and} \quad k = 0,1$$
as
$$\langle \sigma_3 \rangle_ k = (-)^k \frac{1-\bar z z}{1+\bar z z},\qquad  \langle \sigma_1 \rangle_ k = (-)^k  \frac{\bar z + z}{1+\bar z z}
,\qquad  \langle \sigma_2 \rangle_ k = i (-)^k  \frac{\bar z -
z}{1+\bar z z}$$ where $ z = - i \frac{\eta}{\sqrt{\bar\eta \eta}}
\tan \sqrt{\bar\eta \eta} $. They can be also written as
$$\langle \sigma_3 \rangle_ k = (-)^k \cos \theta,\qquad  \langle \sigma_1 \rangle_ k = (-)^k  \sin \theta
\cos \varphi,  \qquad  \langle \sigma_2 \rangle_ k = (-)^k \sin
\theta \sin \varphi$$ where $\frac{\theta}{2}e^{i\varphi} = -i
\eta$. These mean values combined with the expressions of the
correlation matrix elements (\ref{R03+R30}) and (\ref{Rii})
determine the explicit form of the conditional densities
(\ref{rhoBk}) and the conditional entropy (\ref{condit-entropy}) in
terms of the angular variables $\theta$ and $\varphi$. Indeed,
combining the equations (\ref{rhoBk}), (\ref{Eq:VNmsur}),
(\ref{Xform}) and (\ref{operatorU}), it is simply seen that the
density operators $\rho_k^B$ take the following form
\begin{eqnarray}
\rho_k^B = \frac{1}{p_k^B}\left(%
\begin{array}{cc}
  (1+ R_{03}) + (R_{30}+ R_{33})\langle \sigma_3\rangle_k & R_{11}\langle \sigma_1\rangle_k - i R_{22}\langle \sigma_2\rangle_k\\
 R_{11}\langle \sigma_1\rangle_k + i R_{22}\langle \sigma_2\rangle_k & (1- R_{03}) + (R_{30}- R_{33})\langle \sigma_3\rangle_k \\
\end{array}%
\right)
\end{eqnarray}
where
$$p_k^B = \frac{1}{2} ( 1 + R_{30}~\langle \sigma_3\rangle_k).$$
It follows that the conditional entropy  given by
(\ref{condit-entropy}) rewrites also as
\begin{eqnarray}
\widetilde{S} \equiv \widetilde{S}(\theta, \varphi) = \sum_{k = 0 ,
1} p_k^B ~ H \bigg(\frac{1}{2} + \frac{1}{2} \sqrt{1 - 4\det
\rho_k^B}\bigg)\label{stilde}
\end{eqnarray}
and can explicitly expressed as a function of $\theta$ and
$\varphi$. Then, the minimization of $\widetilde{S}$ can performed
over the polar and azimuthal angles. Nevertheless, there exists
another elegant way to optimize the conditional entropy. It is based
on the Koashi-Winter relation \cite{Koachi-Winter} (see also
\cite{Shi1}) as we shall explain in what follows.
\subsection{Minimization of conditional entropy}
The Koashi-Winter relation establishes the connection between the
classical correlation of a bipartite state $\rho^{AB}$ and the
entanglement of formation of its complement $\rho^{BC}$. This
connection requires the purification of the state $\rho^{AB}$. In
this respect, as $\rho^{AB}$ is a two-qubit state of rank two, it
decomposes as
\begin{eqnarray}
\rho^{AB} = \lambda_1 \vert \psi_1 \rangle \langle \psi_1 \vert +
\lambda_2 \vert \psi_2 \rangle \langle \psi_2 \vert
\end{eqnarray}
where the eigenvalues $\lambda_1$ and $\lambda_2$ are given by
(\ref{lambda12}) and the eigenstates $\vert \psi_1 \rangle$ and
$\vert \psi_2 \rangle$ are
\begin{eqnarray}
\vert \psi_1 \rangle = \frac{1} {\sqrt{a^2_{\alpha}a^2_{\alpha t} +
b^2_{\alpha}b^2_{\alpha t}}} ( a_{\alpha}a_{\alpha t} \vert
u_{\alpha},u_{\alpha t} \rangle + b_{\alpha}b_{\alpha t} \vert
v_{\alpha},v_{\alpha t} \rangle)
 \nonumber \\
\vert \psi_2 \rangle = \frac{1} {\sqrt{a^2_{\alpha}b^2_{\alpha t} +
b^2_{\alpha}a^2_{\alpha t}}} ( a_{\alpha}b_{\alpha t} \vert
u_{\alpha}, v_{\alpha t} \rangle + b_{\alpha}a_{\alpha t} \vert
v_{\alpha}, v_{\alpha t} \rangle).
\end{eqnarray}
Attaching a qubit $C$ to the bipartite system $AB$, we write the
purification of $\rho^{AB}$ as
\begin{eqnarray}
\vert \psi \rangle = \sqrt{\lambda_1} \vert \psi_1 \rangle \otimes
\vert u_{\alpha} \rangle +  \sqrt{\lambda_2} \vert \psi_2 \rangle
\otimes \vert v_{\alpha} \rangle
\end{eqnarray}
such that the whole system $ABC$ is described by the pure density
state $\rho^{ABC} = \vert \psi \rangle \langle \psi \vert $ from
which one has  the bipartite densities $\rho^{AB} = {\rm Tr}_C
\rho^{ABC}$ and $\rho^{BC} = {\rm Tr}_A \rho^{ABC}$. Suppose now
that a von Neumann measurement $\{ M_0 , M_1\}$ is performed on
qubit $A$ (here also we need positive operator valued measurement of
rank one that is proportional the one dimensional projector). From
the viewpoint of the whole system in the pure state $\vert \psi
\rangle$, the measurement gives rise to an ensemble for $\rho^{BC}$
that we denote by
$${\cal E}^{BC} = \{ p_k , \vert \phi_k^{BC} \rangle \}$$
where
$$ p_k = \langle \psi \vert M_k \otimes \mathbb{I} \otimes \mathbb{I} \vert \psi \rangle \qquad
 \vert \phi_k^{BC} \rangle \langle \phi_k^{BC} \vert =   \frac{1}{p_k} {\rm Tr}_A
\bigg[ (M_k \otimes \mathbb{I} \otimes \mathbb{I}) \vert \psi
\rangle \langle \psi \vert \bigg].$$ On the other hand, from the
viewpoint of the state $\rho^{AB}$, the von Neuman measurement on
$A$ gives rise to the ensemble for $\rho^B$  defined previously as
${\cal E}^{B} = \{ p^B_k , \rho_k^{B} \}$. It is simple to check
that the ensemble ${\cal E}^{B} $ can be induced from ${\cal
E}^{BC}$ by tracing out the qubit $C$, namely
$$\rho_k^{B} = {\rm Tr}_C \bigg[\vert \phi_k^{BC}  \rangle \langle \phi_k^{BC}  \vert \bigg].$$
We denote by $E(\vert \phi_k^{BC}  \rangle )$ the measure of
entanglement for pure states. It is given by the von Neumann entropy
of the reduced subsystem $ \rho_k^{B} = {\rm Tr}_C (\vert
\phi_k^{BC} \rangle \langle \phi_k^{BC} \vert)$
$$ E (\vert \phi_k^{BC}  \rangle ) = S (\rho_k^{B}).$$
It follows that the average of entanglement of formation over the
ensemble ${\cal E}^{BC}$
$$\overline{E}^{BC} = \sum_{k = 0, 1} p_k  E (\vert \phi_k^{BC}  \rangle )$$
coincides with the conditional entropy  (\ref{condit-entropy}).At
this level, it is important to notice that Koachi and Winter
\cite{Koachi-Winter} have pointed out that the minimum value of
$\overline{E}^{BC}$ is exactly the entanglement of formation of
$\rho^{BC}$. Consequently,
\begin{equation}
 \widetilde{S}_{\rm min} = E(\rho^{BC}).
\end{equation}
This relation simplifies drastically the minimization process of the
conditional entropy. Therefore, employing the prescription presented
in \cite{Wootters-PRL80-1998} to get the concurrence of the density
$\rho^{BC}$,  one obtains
\begin{equation}
\widetilde{S}_{\rm min} = E(\rho^{BC}) = H\bigg(\frac{1}{2} +
\frac{1}{2} \sqrt{1 - \vert C(\rho^{BC})\vert^2}\bigg)\label{Smin}
\end{equation}
with
$$ C(\rho^{BC}) = \frac{\sqrt{p^{2}(1 - p^{2r^2})(1 - p^{2t^2})}}{(1+p^{2})}.$$
It is remarkable that the minimal value of the conditional entropy
given by (\ref{Smin}) is reached for $\theta = \pi/2$ and $\varphi =
0$. Indeed, using the explicit form of the conditional entropy
(\ref{stilde}) in term of the polar and azimuthal angles $\theta$
and $\phi$, one can verify  that
$$ \widetilde{S}_{\rm min} =  \widetilde{S} (\theta = \frac{\pi}{2}, \phi = 0).$$
It is important to stress that the special values $\theta = 0$ and
$\phi = \pi/2$ coincide with ones obtained in \cite{Ali} (see also
\cite{X-M Lu}) for optimal  measurements to access quantum discord
of two-qubit states. According to the equation (\ref{def: classical
correlation}), the classical correlation is
\begin{equation}
C(\rho^{AB}) = H\bigg( \frac{1}{2} + \frac{1}{2} \frac{p^{t^2} +
p^{r^2+1}}{1+p^2}\bigg) - H\bigg( \frac{1}{2} + \frac{1}{2}
\frac{\sqrt{1+p^2 + p^{2r^2+2} + p^{2t^2+2}}}{1+p^2}\bigg)
\end{equation}
and using the definition (\ref{def: discord}), the explicit
expression of quantum discord reads
\begin{equation}\label{qd-formula}
D(\rho^{AB}) = H\bigg( \frac{1}{2} + \frac{p}{1+p^2}\bigg) + H\bigg(
\frac{1}{2} + \frac{1}{2} \frac{\sqrt{1+p^2 + p^{2r^2+2} +
p^{2t^2+2}}}{1+p^2}\bigg) - H\bigg( \frac{1}{2} + \frac{1}{2}
\frac{p^{r^2} + p^{t^2+1}}{1+p^2}\bigg).
\end{equation}
Note that for $r = 0$, the density state $\rho^{AB}$
(\ref{density-main}) reduces to the pure density Bell cat-states
(\ref{bell}) and the  the quantum discord (\ref{qd-formula}) gives
\begin{equation}
D(\vert B_{\alpha,\alpha}\rangle \langle B_{\alpha,\alpha} \vert) =
H \bigg(\frac{1}{2} + \frac{p}{1 + p^2}\bigg)
\end{equation}
which is, as expected, exactly the entanglement of formation $E
(\vert B_{\alpha,\alpha}\rangle \langle B_{\alpha,\alpha}\vert)$.
Indeed, it is simple to check that the concurrence of the pure Bell
cat-state $\vert B_{\alpha,\alpha}\rangle $ is
\begin{equation}
C(\vert B_{\alpha,\alpha}\rangle) = \frac{1 - p^2}{1 + p^2}
\end{equation}
which coincides for $r = 0$ with the concurrence of the state
(\ref{density-main}) given by
\begin{equation}
C (\rho^{AB})=  \frac{p^{r^2}\sqrt{(1 - p^2)(1 - p^{2t^2})}}{1 +
p^2}.\label{crhoAB}
\end{equation}
where we used the definition of the concurrence for mixed states
introduced in \cite{Wootters-PRL80-1998}. It is also simply verified
from (\ref{qd-formula}) that for $r = 1$, the quantum discord
vanishes and the concurrence (\ref{crhoAB}) is also zero.

The behavior of quantum discord (\ref{qd-formula}) versus the
overlapping $p$ and the reflection coefficient $r$ is plotted in the
figure 1. As seen from this figure, for $p$ fixed the quantum
discord decreases as we increase the reflection parameter $r$ of the
beam splitter and the maximum is obtained in the limiting case $r =
0$. This indicates that the noisy channel, inducing the decoherence
effects, renders the system less correlated and subsequently the
quantum discord decreases. In other hand, for a fixed value of $r$,
the quantum discord starts increasing, reaches a maximal value and
decreases after. To see this behavior, we plot in the figure 2, the
quantum discord versus the overlapping for Bell cat-states for
different values of the reflection coefficient $r$. In particular,
under the action of a 50:50 beam splitter, the maximal value of
quantum discord is reached for $p \simeq 0.4$. Clearly, for $r = 0$,
the maximum of the quantum discord or entanglement of formation is
attained for $p = 1$. From the figure 2, it is easily seen that the
value of the overlap $p$, for which the quantum discord $D$ is
maximal, increases for increasing values of $r$.

\begin{center}
\includegraphics[width=4in]{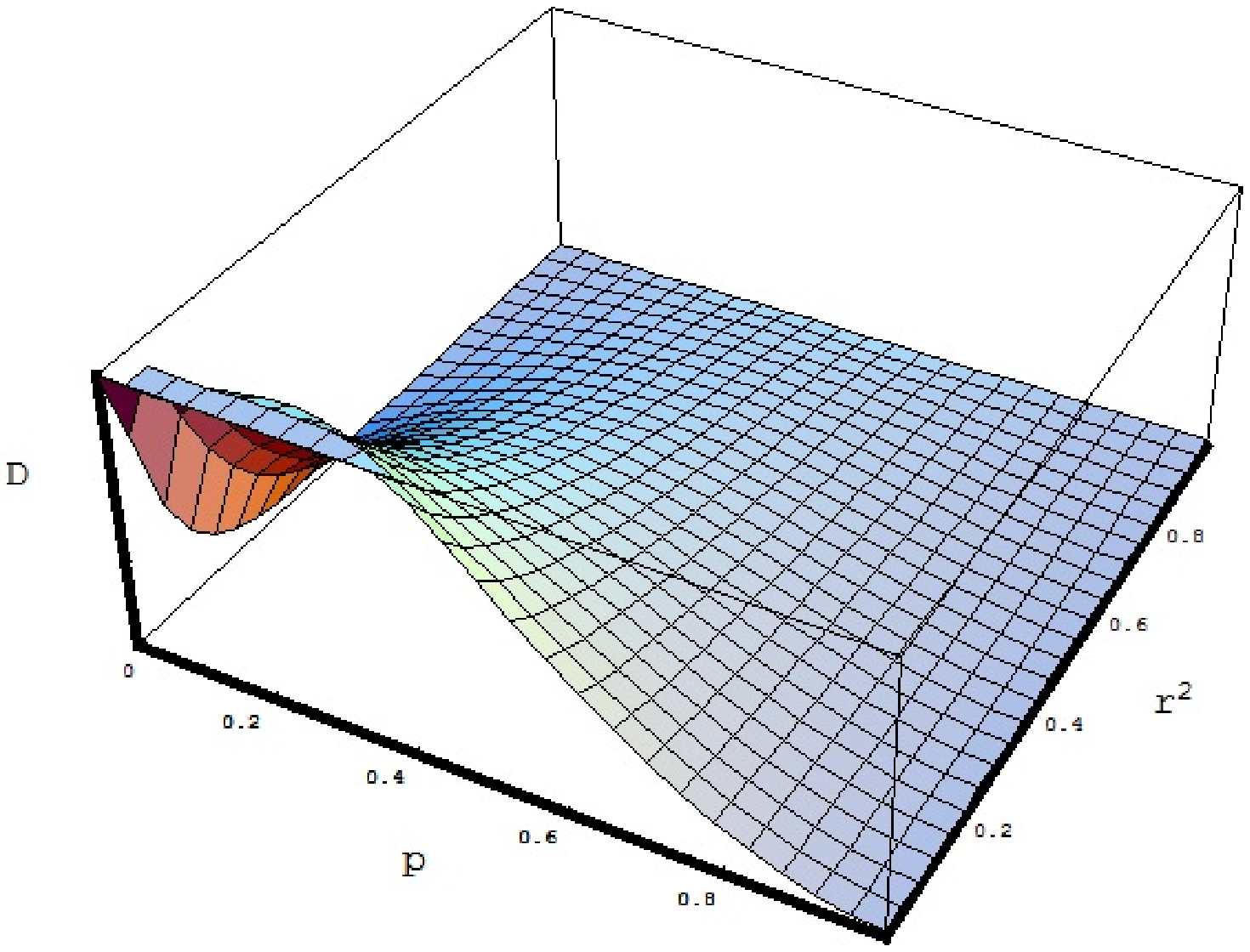}\\
FIG. 1:  {\sf The pairwise quantum discord $D$ versus the
overlapping $p$ of Glauber states and the reflection parameter
$r^2$.}
\end{center}
\begin{center}
\includegraphics[width=4in]{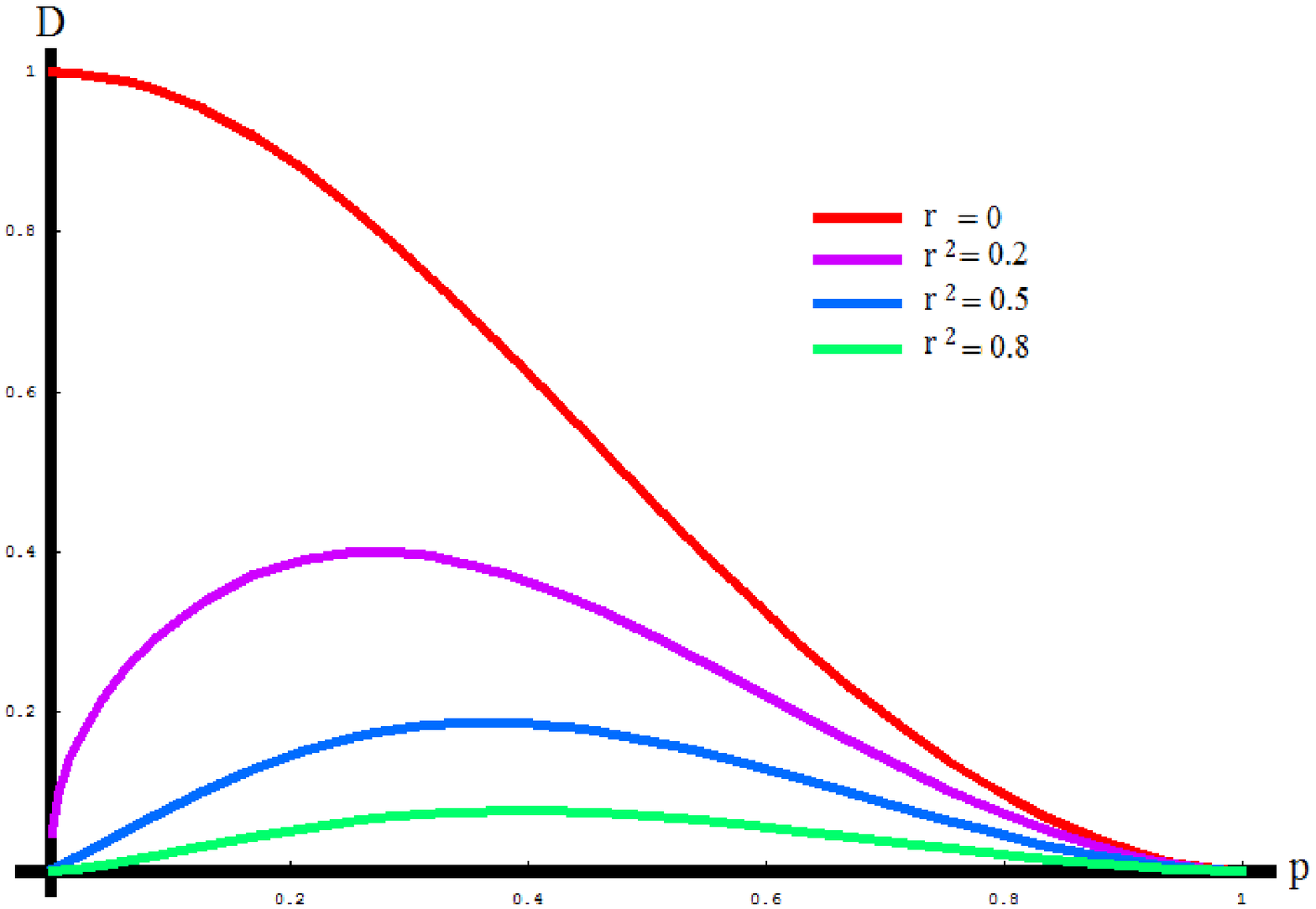}\\
FIG. 2:  {\sf The pairwise quantum discord $D$ versus the
overlapping $p$ for different values of $r$}
\end{center}

\section{Evolution of quantum correlations under dephasing channel }

In this section, we investigate the dynamics of bipartite quantum
correlations (entanglement and quantum discord) of  damped Bell
cat-states, given by the density $\rho^{AB}$ (\ref{Xform}), under
dephasing channel. To describe conveniently the effect  of this
channel, we use the the Kraus operator approach (see for instance
\cite{NC-QIQC-2000}). In this approach, the time evolution of the
bipartite density $\rho^{AB} \equiv \rho^{AB}(0)$ (\ref{Xform}) can
be written compactly as
$$ \rho^{AB}(t) = \sum_{\mu, \nu}E_{\mu , \nu}(t) ~\rho^{AB}(0)~ E_{\mu , \nu}^{\dagger}(t) $$
where the so-called Kraus operators
$$E_{\mu , \nu}(t) = E_{\mu}(t)\otimes E_{\nu}(t) \qquad \sum_{\mu,
\nu}E_{\mu , \nu}^{\dagger}  E_{\mu , \nu} = \mathbb{I}.$$ The
operators $E_{\mu}$ describe the one-qubit quantum channel effects.
For a dephasing channel,  the non-zero Kraus operators  are given by
$$ E_0 = {\rm diag}( 1 , \sqrt{1 - \gamma})  \qquad  E_1 = {\rm diag}( 1 , \sqrt{\gamma})$$
with $ \gamma = 1 - e^{-\Gamma t}$ and $\Gamma$ denoting the decay
rate. This gives
\begin{eqnarray}
\rho^{AB}(t) =  \frac{2}{N_{\alpha}}\left(
\begin{array}{cccc}
(1+c)a^2_{\alpha}a^2_{\alpha t} & 0 & 0 & e^{-\Gamma
t}(1+c)a_{\alpha}a_{\alpha t}b_{\alpha
}b_{\alpha t}\\
0 & (1-c)a^2_{\alpha }b^2_{\alpha t} & e^{-\Gamma
t}(1-c)a_{\alpha}a_{\alpha t}b_{\alpha
}b_{\alpha t} & 0 \\
0 & e^{-\Gamma t}(1-c)a_{\alpha }a_{\alpha t}b_{\alpha }b_{\alpha t}
& (1-c)b^2_{\alpha }a^2_{\alpha t}
& 0 \\
e^{-\Gamma t}(1+c)a_{\alpha} a_{\alpha t}b_{\alpha }b_{\alpha t} & 0
& 0 & (1+c)b^2_{\alpha }b^2_{\alpha t}
\end{array}
\right). \label{Xstate(t)}
\end{eqnarray}
The entanglement in this state is measured by the concurrence
$$ C = 2~ {\rm max} \{ 0 , \Lambda_1(t) , \Lambda_2(t) \}$$
where
$$ \Lambda_1(t) =  \frac{2}{N_{\alpha}}a_{\alpha }a_{\alpha t}b_{\alpha }b_{\alpha
t}\bigg[ (1 - \gamma) (1 + c) - (1-c)\bigg]
 \qquad  \Lambda_2(t) = \frac{2}{N_{\alpha}}a_{\alpha }a_{\alpha t}b_{\alpha }b_{\alpha t}\bigg[ (1 - \gamma) (1 - c) - (1+c)\bigg].$$
Since $\Lambda_2(t)$ is non positive, the concurrence is
\begin{equation}
C (t)= \frac{1}{2} \frac{\sqrt{(1 - p^2)(1 - p^{2t^2})}}{1 +
p^2}\bigg[ e^{-\Gamma t}(1 + p^{r^2}) -
(1-p^{r^2})\bigg]\label{c(t)}
\end{equation}
for
$$ t < t_0 = \frac{1}{\Gamma} [ ~\ln(1 + p^{r^2}) -  \ln(1 - p^{r^2})].$$
In this case, the system is entangled. However, for $ t \ge t_0$,
the concurrence is zero and the entanglement disappears., i.e. the
system is separable. This shows clearly that under dephasing
channel, the entanglement suddenly vanishes. Note that the bipartite
system under consideration is entangled in the absence of external
noise. Indeed, for $ t = 0$, the concurrence (\ref{c(t)}) reproduces
(\ref{crhoAB}) which is non zero except in the limiting case $ p
\longrightarrow 0$ ( $\vert \alpha \vert \longrightarrow + \infty$)
or $ r = 1$. The phenomenon of total loss of entanglement, termed in
the literature "entanglement sudden death" \cite{Yu}, was
experimentally confirmed under some specific conditions
\cite{Almeida}. As the concurrence vanishes after a finite time
$t_0$, it is interesting to ask what happens to quantum discord. The
explicit expression of the amount quantum discord in the state
$\rho^{AB}(t)$ can be obtained following the method described in the
previous section. But, here we need only to know if the state
$\rho^{AB}(t)$ has vanishing quantum discord. As the density
$\rho^{AB}(t)$ has also the form of the letter $X$, one can use the
criteria, classifying the so-called  $X$ states with vanishing
quantum discord, discussed in \cite{Bylicka}. Therefore, according
to this criteria, $\rho^{AB}(t)$ has zero quantum discord if and
only if $ p \longrightarrow 0$. The quantum discord in the state
$\rho^{AB}(t)$ is in general nonzero even when entanglement suddenly
disappears. This gives a special instance of separable quantum
states for which the quantum discord is non zero. This agrees with
the commonly accepted fact that the quantum discord is a kind of
quantum correlations which goes beyond entanglement and almost all
quantum states have non vanishing quantum correlations
\cite{Ferraro}.

\section{Concluding remarks}
To close this paper, let us briefly summarize the main results. Our
effort was devoted to investigate the evolution of quantum discord
of Bell cat-states under amplitude damping channel. We have derived
the analytical expressions for the classical correlation and quantum
discord. We discussed the usefulness of the Koashi-Winter relation
to determine a closed form of quantum discord present in the system.
We also discussed the evolution of the transmitted Bell cat-states
under a dephasing channel. This provides a special instance for
separable mixed quantum states with non vanishing quantum discord.
The method used  in this paper, to derive quantum discord,
constitutes an alternative way to compute analytically classical
correlations and quantum discord. It can be applied easily in
evaluating the quantum discord present in the other Bell cat-states:
$$\vert B^-_{\alpha, \alpha}\rangle \sim \vert
\alpha, \alpha \rangle -  \vert - \alpha, - \alpha \rangle$$
$$\vert B^+_{\alpha, -\alpha}\rangle \sim \vert
\alpha, -\alpha \rangle +  \vert - \alpha,  \alpha \rangle$$
$$\vert B^-_{\alpha, -\alpha}\rangle \sim \vert
\alpha, -\alpha \rangle -  \vert - \alpha,  \alpha \rangle$$ after
passing trough a amplitude damping channel. In addition, the
analysis presented here for the evaluation of quantum discord is
readily extended to more general systems, including  squeezed
states, $SU(2)$ and $SU(1,1)$ coherent states and so on. Finally, it
will be interesting to compare the quantum discord in bipartite
coherent states with its geometrized version usually called in the
literature geometric quantum discord \cite{Dakic2010}. Further
thought in this direction might be worthwhile.

\end{document}